\documentclass[
5p
]{elsarticle}

\usepackage{lineno}
\usepackage[dvipdfmx]{hyperref}
\modulolinenumbers[5]

\journal{Journal of \LaTeX\ Templates}

\usepackage{amsmath,amssymb}
\usepackage{graphicx}
\usepackage{dcolumn}
\usepackage{bm}
\usepackage{color}
\biboptions{sort&compress}
\usepackage[normalem]{ulem}  
\def\br#1{\left( #1 \right)}
\def\bra#1{\left\{ #1 \right\}}
\def\brac#1{\left[ #1 \right]}

\def\non{\nonumber}

\begin{document}

\begin{frontmatter}

\title{
Search of QCD phase transition points in the canonical approach of the NJL model 
}

\author[ad1,ad2,ad3]{Masayuki~Wakayama}
\author[ad1,ad4]{Atsushi~Hosaka}
\address[ad1]{Research Center for Nuclear Physics (RCNP), Osaka University, Ibaraki, Osaka 567-0047, Japan}
\address[ad2]{Department of Physics, Pukyong National University (PKNU), Busan 48513, Republic of Korea}
\address[ad3]{Center for Extreme Nuclear Matters (CENuM), Korea University, Seoul 02841, Republic of Korea}
\address[ad4]{Advanced Science Research Center, Japan Atomic Energy Agency (JAEA), Tokai 319-1195, Japan}

\begin{abstract}

We study the Lee-Yang zeros in the canonical approach to search phase transition points at finite temperature and density 
in the Nambu-Jona-Lasinio (NJL) model as an effective model of QCD. 
The canonical approach is a promising method to avoid the sign problem in lattice QCD at finite density. 
We find that a set of Lee-Yang zeros computed with finite degrees of freedom can be extrapolated to those with infinite degrees of freedom, 
providing the correct phase transition point. 
We propose the present method as a useful method for actual lattice simulations for QCD. 

\end{abstract}


\end{frontmatter}


\section{\label{intro}Introduction}

The role of Quantum Chromodynamics (QCD), the theory of the strong interaction, at finite temperature and density 
is increasing 
as it provides basic inputs in the fundamental questions 
such as 
the matter generation in the early universe, formation of galaxies and stars, and mysterious stellar objects such as neutron stars and black-halls. 
Especially, the latter objects are under active discussions 
due to the recent observation of gravitational waves~\cite{Abbott:2016blz,TheLIGOScientific:2017qsa} and black-halls~\cite{Akiyama:2019eap}. 
Experimentally, those problems are approached by the high energy accelerators at such as 
J-PARC (KEK/JAEA), FAIR (GSI) and NICA (JINR), which will be expected to operate in the near future. 
Theoretically, lattice QCD is known as a 
unique method 
for the first principle calculations of QCD. 

However, lattice QCD suffers from the sign problem at finite density. 
Many methods have been proposed toward avoiding the sign problem. 
Meanwhile, a method called the canonical approach~\cite{Hasenfratz:1991ax} 
has been recently developed rapidly 
with multiple-precision arithmetic~\cite{Morita:2012kt,Fukuda:2015mva,Nakamura:2015jra,deForcrand:2006ec,
Ejiri:2008xt,Li:2010qf,Li:2011ee,Danzer:2012vw,Gattringer:2014hra,
Boyda:2017lps,Goy:2016egl,Bornyakov:2016wld,Boyda:2017dyo,Wakayama:2018wkc}. 
In the canonical approach physical quantities are calculated at pure imaginary chemical potentials, in which the sign problem does not exist. 
Information at physical real finite chemical potentials is extracted. 
The canonical approach can predict physical observable such as 
particle distributions in heavy ion collisions and reveal the phase structure at high densities. 

In Ref.~\cite{Wakayama:2018wkc}, 
one of the present authors studied the so-called Lee-Yang zeros (LYZs), 
which are zeros of grand canonical partition functions as functions of the fugacity parameter. 
LYZs provide us with various information of phase transitions~\cite{Yang:1952be,Lee:1952ig}. 
However, numerical simulations can be only available with finite degrees of freedom, 
which should be extrapolated to the real situation with infinite degrees of freedom. 
Currently, such extrapolation method is not known.

To attack this problem, we propose to study in an effective model of QCD, the Nambu-Jona-Lasinio (NJL) model~\cite{Nambu:1961tp,Nambu:1961fr}. 
Because phase properties are known well in the model e.g. \cite{Kunihiro:1991qu,Hatsuda:1994pi}, 
we can study exclusively the effect of finite degrees of freedom. 
We show how the LYZs with finite degrees of freedom are smoothly extrapolated to those with infinite degrees of freedom. 
Moreover, we also study an additional approximation for the number density as a function of imaginary chemical potential 
which has been used in the lattice simulations~\cite{Boyda:2017lps,Goy:2016egl,Bornyakov:2016wld,Boyda:2017dyo,Wakayama:2018wkc}. 
Taking into account the two features, phase transition points are well determined from the data of finite degrees of freedom.  

We will begin, in Sec.~\ref{cano}, by briefly describing the canonical approach in the NJL model. 
In Sec.~\ref{Z_LYZ}, we introduce LYZs and a computational method of LYZs. 
In Sec.~\ref{simu}, the numerical results are shown. 
After that, we discuss the extrapolation procedure 
from the analysis of finite degrees of freedom to the case of infinite degrees of freedom. 
Section~\ref{summ} is devoted to the summary.

\section{\label{cano}Canonical approach in the NJL model}

Let us begin with a brief review of the canonical approach. 
The grand canonical partition function $Z_{GC}$ 
at a quark chemical potential $\mu$, a temperature $T$ and a volume $V$ of the system can be written as 
\begin{eqnarray}
Z _{\rm GC} (\mu,T,V) 
&=& \sum_{n=-\infty}^{\infty} \left \langle n \right | e^{-\br{\hat H - \mu \hat N}/T} \left | n \right\rangle \non \\
&=& \sum_{n=-\infty}^{\infty} Z_C(n,T,V) \xi^n \ , \label{fuga_exp}
\end{eqnarray}
where $\hat H$, $\hat N$, $\xi$ and $Z_C(n,T,V)$ are 
the Hamiltonian operator, the quark number operator, the quark fugacity defined by $\xi = e^{\mu/T}$ and the canonical partition functions, respectively. 
Applying Fourier transformation to $Z_{GC}$ at the pure imaginary chemical potential $\mu=i\mu_{I}$ $(\mu_{I}\in \mathbb{R})$, 
we obtain the canonical partition functions,  
\begin{eqnarray}
\!\!\!\!\!\!\!\!
Z_C(n,T,V)  &=& \int_0^{2\pi} \frac{d\theta}{2\pi} \, e^{-in\theta} Z_{\rm GC} (i \mu_{I},T,V) \ , \ \ \ 
\label{ZC}
\end{eqnarray}
where $\theta = \frac{\mu_{I}}{T}$. 
In order to suppress the cancellation of significant digits 
that comes from the high frequency of $e^{-in\theta}$ at large $n$, 
we perform Fourier transformation with multiple-precision arithmetic. 

In lattice QCD calculations, moreover, 
the integration method~\cite{Boyda:2017lps,Goy:2016egl,Bornyakov:2016wld,Boyda:2017dyo,Wakayama:2018wkc} 
is used to extract $Z_C$ for large $n$. 
In the integration method, 
$Z_{GC}(i\mu_{I})$ is evaluated from the number density, 
\begin{eqnarray}
 \frac{n_{q}}{T^3}(i\mu_I) &=& \frac{1}{VT^2} \frac{\partial}{\partial (i\mu_I)} \ln Z_{GC}(i\mu_I) \ . 
\end{eqnarray}
Since $Z_{GC}(i\mu_{I})$ is a real quantity, 
we define $n_q$ by the real valued $n_{qI}$ by $n_{q}=in_{qI}$ at $\mu=i\mu_I$. 
It is well known that 
$n_{qI}$ 
can be approximated by a Fourier series, 
\begin{eqnarray}
 \frac{n_{qI}}{T^3}(\theta) &=& \sum_{k=1}^{N_{{\sin}}} f_{k} \sin(k\theta) \ , 
 \label{mulsin}
\end{eqnarray}
with 
a finite number of terms of $N_{\sin}$~\cite{DElia:2009pdy,Takaishi:2010kc}. 
Fitting the Fourier series to $n_{qI}$, we can evaluate $Z_{GC}$ at the imaginary $\mu$ in good approximation from 
\begin{eqnarray}
\!\!\!\!\!\!\!\!\!\!\!\!
Z_{\rm GC} (i \mu_{I},T,V) 
&=& C \exp{\bra{-V \int^{\theta}_0 d\theta^{\prime} \, n_{qI}(\theta^{\prime})}} \non \\
&=& C \exp{\bra{VT^3 \sum_{k=1}^{N_{\sin}} \frac{f_k}{k} \cos \br{k\theta} }}  \ , \ \ \ \ \ 
\label{GCint}
\end{eqnarray}
where $C$ is an integration constant. 

In this paper, we compute $n_{qI}$ in Eq.~\eqref{mulsin} in the NJL model. 
The Lagrangian density of the NJL model is 
\begin{eqnarray}
\!\!\!\!\!\!\!\!\!\!\!\!\!
\cal{L} &=& \bar{\psi} \br{i \gamma^{\nu} \partial_{\nu} - m_q} \psi 
                + G \brac{\br{\bar{\psi}\psi}^2  + \br{\bar{\psi} i \gamma_5 \vec{\tau} \psi}^2} \, , \ \ \ \ 
\end{eqnarray}
where $\psi$ is the quark field with two flavors ($N_f=2$) and three colors ($N_c=3$)~\cite{Kunihiro:1991qu,Hatsuda:1994pi}. 
Given the current quark mass $m_q=5.5$\ [MeV], 
the remaining parameters, the coupling constant $G$ and the cutoff $\Lambda$ 
are determined such that the pion decay constant $f_{\pi} = 93$\ [MeV] and the constituent quark mass $m_q = 335$\ [MeV] 
are reproduced in the mean field approximation, $G = 5.5$ \ [GeV$^{-2}$] and $\Lambda = 631$\ [MeV]. 
First we investigate the case of infinite volume.  
In this case, 
the thermodynamic potential (per unit volume) $\omega$ 
is defined by utilizing the Matsubara formula as 
\begin{eqnarray}
\!\!\!\!\!\!\!\!\!\!\!\!\!
\omega(T,\mu) && \non \\
     && \!\!\!\!\!\!\!\!\!\!\!\!\!\!\!\!\!\!\!\!\!\!
     = \ \frac{1}{2G} \br{M-m_q}^2 \non \\ 
     && \!\!\!\!\!\!\!\!\!\!\!\!\!\!\!\!\!\!\!\!\!\!
      - \  2N_c N_f \int \frac{d^3p}{\br{2\pi}^3} 
                         \Big\{E_{p} 
                             + T\ln\brac{1+e^{-\br{E_p+\mu}/T}} \non \\
                         && \ \ \ \ \ \ \ \ \ \ \ \ \ \ \ \ \ \ \ \ 
                            + T\ln\brac{1+e^{-\br{E_p-\mu}/T}}\Big\} \, , \ \ \ \ 
\end{eqnarray}
where the energy and the constituent quark mass are given as $E_p = \sqrt{p^2+M^2}$ and $M = m_q - G\sigma$, respectively. 
The chiral condensate $\sigma\equiv\langle \bar{\psi} \psi \rangle$ 
is evaluated from the stationary condition $\partial \omega/ \partial \sigma = 0$. 

Since the number density $n_q$ is defined as 
$n_q \equiv -\partial \omega / \partial \mu$, 
we find in the NJL model 
\begin{eqnarray}
\!\!\!\!\!\!\!\!\!\!\!\!\!
n_{qI}(\theta) &=& \frac{1}{G} \frac{\partial M}{\partial \mu_I}\br{M-m_q} \non \\
   &-& 2N_c N_f \int \frac{d^3p}{\br{2\pi}^3} \Bigg\{\frac{\partial E_p}{\partial \mu_I} \non \\
     && 
     - \frac{\partial E_p}{\partial \mu_I}\frac{2\br{1+e^{E_p/T}\cos\theta}}{\br{1+e^{E_p/T}\cos\theta}^2+e^{2E_p/T}\sin^2\theta}  \non \\
     && 
     - \frac{2e^{E_p/T}\sin\theta}{\br{1+e^{E_p/T}\cos\theta}^2+e^{2E_p/T}\sin^2\theta}  
     \Bigg\} \, . \label{nqI_NJL}
\end{eqnarray}
Note that the constituent quark mass $M$ in Eq.~\eqref{nqI_NJL} obeys the following gap equation: 
\begin{eqnarray}
\!\!\!\!\!\!\!\!\!\!\!\!\!\!\!
 M  \!\!\!\!&=&\!\!\!\! m_q + 2N_c N_f GM  \non \\
\!\!\!\!\!\!\!\!\!\!\!\!\!\!\!
      \!\!&\times&\!\!\!\!\!\!     \int \!\!\frac{d^3p}{\br{2\pi}^3} \frac{1}{E_p} \Bigg\{1  - \frac{2\br{1+e^{E_p/T}\cos\theta}}{\br{1+e^{E_p/T}\cos\theta}^2+e^{2E_p/T}\sin^2\theta}  \Bigg\}  . \! \non \\
\!\!\!\!\!\!\!\!\!\!\!\!\!\!\!
      && \!\!
\end{eqnarray}

\begin{table*}
\caption{
Results for fitting coefficients $f_{k}$ from the data of $n_{qI}/T^3$ for each temperature. 
The values of $f_k$ in the list are rounded to three significant digits 
although errors for all $f_k$ are very small, 
$\delta f_k < 3\times 10^{-15}$.}
\begin{tabular}{c|ccccccc}\hline \hline
 $T$\ [MeV] & $f_1$ & $f_2$ & $f_3$ & $f_4$  & $f_5$  & $f_6$  & $f_7$ \\\hline 
29 & $6.51\times10^{-4}$ & $1.87\times10^{-9}$ & --- & --- & --- & ---  & --- \\
39 & $8.56\times10^{-3}$ & $6.15\times10^{-7}$ & $6.52\times10^{-11}$  & --- & --- & --- & --- \\
49 & $3.69\times10^{-2}$ & $1.81\times10^{-5}$ & $1.31\times10^{-8}\, \,$  & $1.11\times10^{-11}$ & --- & --- &  --- \\
59 & $9.31\times10^{-2}$ & $1.63\times10^{-4}$ & $4.24\times10^{-7}\, \,$  & $1.29\times10^{-9}\, \,$ & $4.30\times10^{-12}$ & --- & ---  \\
79 & $2.69\times10^{-1}$ & $2.28\times10^{-3}$ & $2.86\times10^{-5}\, \,$  & $4.19\times10^{-7}\, \,$ & $6.65\times10^{-9}\, \,$ & $1.11\times10^{-10}$  & $1.91\times10^{-12}$ \\ \hline \hline
\end{tabular}
\label{table_coef}
\end{table*}

\section{\label{Z_LYZ}Lee-Yang zeros}

In numerical calculations, 
the fugacity expansion of the grand canonical partition function in Eq.~\eqref{fuga_exp} is truncated at a finite value $N_{\rm max}$ as 
\begin{eqnarray}
\!\!\!\!\!\!\!
Z _{\rm GC} (\mu,T,V) &=&  \sum_{n=-N_{\rm max}}^{N_{\rm max}}  Z_{C}(n,T,V) \xi^n \ . 
\label{Zbaryon}
\end{eqnarray}
Since $N_{\rm max}$ is the maximal value of the net-quark number in the system, 
degrees of freedom of the system are limited by the finite $N_{\rm max}$. 
Therefore, we need to take the limit $N_{\rm max} \to \infty$ 
because a system with a finite degree of freedom does not have a phase transition in the real finite chemical potential. 

The theorems of Yang and Lee~\cite{Yang:1952be,Lee:1952ig} 
are of universal and powerful use to investigate phase structures for a system with finite degrees of freedom. 
The so-called Lee-Yang zeros (LYZs), which are the zeros of grand canonical partition functions in complex fugacity plane, 
provide us with various information of phase transitions. 
LYZs are given as roots of the polynomial equation of degree of $2N_{\rm max}$, 
\begin{eqnarray}
  \xi^{N_{\rm max}}  \sum_{n=-N_{\rm max}}^{N_{\rm max}}  Z_{C}(n,T,V) \xi^n   &=& 0 \ , 
\label{LYeq1}
\end{eqnarray}
in the complex $\xi$ plane. 
As $N_{\rm max}$ increases, a distribution of LYZs forms one-dimensional curves in the complex $\xi$ plane. 
If there is a point where LYZs accumulate and are stable as $N_{\rm max}$ increases, 
the point represents the phase transition point. 
Therefore, we investigate the $N_{\rm max}$ dependence of distributions of LYZs 
near the real positive axis of $\xi$. 

A polynomial equation of high degree of $2N_{\rm max}$ such as Eq.~\eqref{LYeq1} is known as an ill-posed problem 
because of significant cancelations. 
A new method to overcome the difficulty was proposed in Ref.~\cite{Nakamura:2013ska}: 
the cut Baum-Kuchen (cBK) algorithm with multiple-precision arithmetic. 
In order to solve the equation, 
we implement the cBK algorithm and a multiple-precision arithmetic package, FMLIB~\cite{FMLIB}.

\begin{figure}
\begin{center}
\includegraphics[scale=0.45]{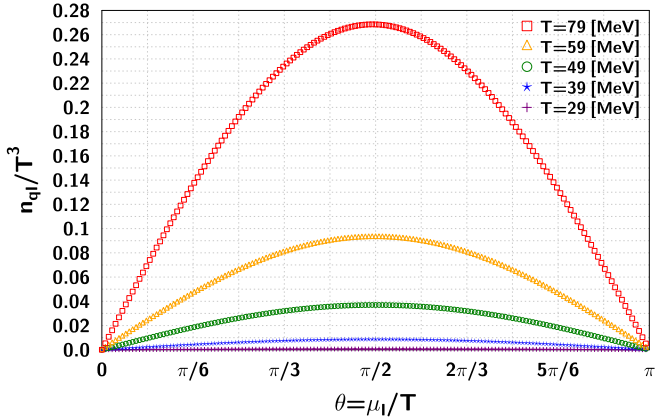}
\caption{ (color online). 
The $\theta$ dependence of the imaginary number density. 
}
\label{thetadep}
\end{center}
\end{figure}

Due to the following properties of $Z_C$, it is sufficient to search 
the LYZs only inside the upper half of the unit circle in the complex $\xi$ plane, 
(i) $Z_C$ satisfy 
\begin{eqnarray}
 Z_C(+n,T,V) &=& Z_C(-n,T,V) \ ,
\label{Zc+-}
\end{eqnarray}
which comes form the charge-parity invariance of quarks, 
and (ii) $Z_C$ are real. 
If there is a LYZ at $\xi$, LYZs also exist at $\xi^{-1}$ and $\xi^{\ast}$ 
because of the properties (i) and (ii), respectively. 
In this paper, we only show LYZs inside the upper half of the unit circle in the complex $\xi$ plane
since whole LYZs can be reconstructed from them.

\section{\label{simu} Numerical results}

\begin{figure}
\begin{center}
\includegraphics[scale=0.45]{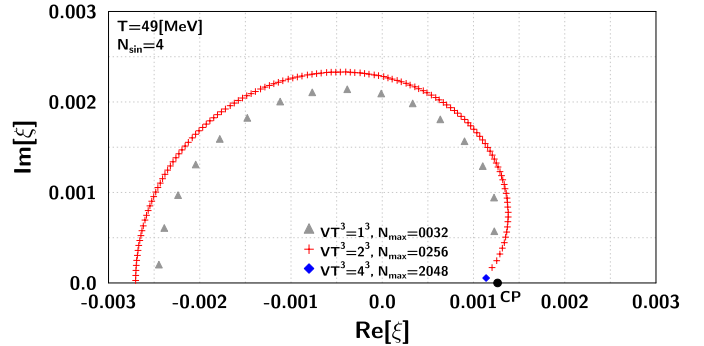}
\caption{ (color online). 
The $V$ dependence of LYZs in the complex $\xi$ plane. 
The black filled circle corresponds to the expected critical point (CP) calculated at the real finite chemical potential. 
}
\label{Vdep}
\end{center}
\end{figure}

\subsection{\label{nqi} Imaginary number density}

Figure~\ref{thetadep} shows the pure imaginary chemical potential dependence of the imaginary number density. 
We calculate the number density $n_{qI}/T^3$ at 161 values of $\mu_{I}$ for various temperature. 
Since there is an anti-symmetry, $n_{qI}(\pi+\theta)=-n_{qI}(\pi-\theta)$, the figure only shows the region $0 \le \theta \le \pi$. 
Because the critical point in the NJL model is already known as $(T_{\rm CP},\mu_{\rm CP})\simeq(49,327)$ [MeV], 
we choose the temperatures at $T_{\rm CP}$ (49 [MeV]), below $T_{\rm CP}$ (29 and 39 [MeV]) and above $T_{\rm CP}$ (59 and 79 [MeV]). 
It turns out that $n_{qI}$ is well fitted by the Fourier series 
of Eq.~\eqref{mulsin} with the coefficients $f_k$ as listed in Table~\ref{thetadep}.

\subsection{\label{nv} Distributions of LYZs}

We calculate the canonical partition functions from the imaginary number density 
by performing Fourier transforms in Eq.~\eqref{ZC} with 5,000 significant digits in decimal notation. 
After the cBK algorithm is carried out with 300 significant digits, 
we obtain LYZs. 
Because, in the cBK algorithm, 
coordinates of LYZs are identified with finite sizes of the annulus sectors,
LYZs have the systematic errors of the cBK algorithm: 
$\delta |\xi|/|\xi| < 1.6\times10^{-2}$ and 
$\delta \br{{\rm arg}\br{\xi}}/{\rm arg}\br{\xi} < 6.5\times10^{-2}$ for all LYZs. 
The systematic errors are tiny 
compared to values of LYZs. 
Therefore, the systematic errors are not displayed in the following figures. 

In Fig.~\ref{Vdep}, we present the distributions of LYZs. 
LYZs form an one-dimensional curve like a circle in the complex $\xi$ plane. 
The black filled point labeled by CP corresponds to the expected critical point evaluated from calculations at real chemical potential.

\begin{figure}
\begin{center}
\includegraphics[scale=0.45]{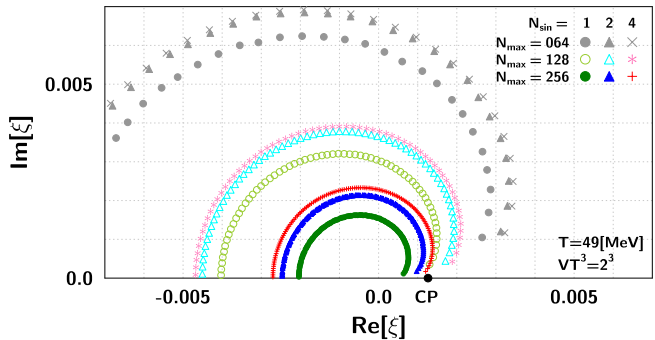}
\caption{ (color online). 
The $N_{\rm max}$ and $N_{\sin}$ dependences of LYZs in the complex $\xi$ plane. 
The black filled circle corresponds to the expected critical point (CP) calculated at the real finite chemical potential. 
}
\label{Nsindep}
\end{center}
\end{figure}

\begin{figure}
\begin{center}
\includegraphics[scale=0.55]{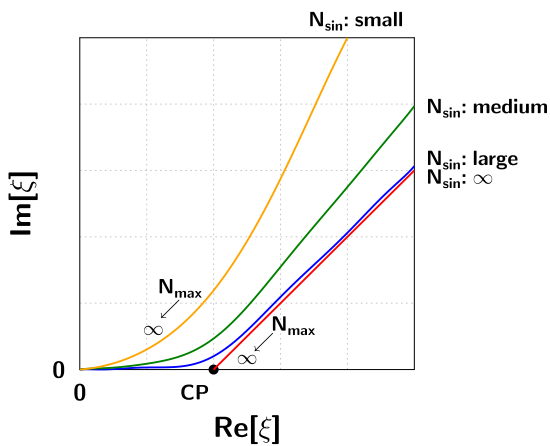}
\caption{ (color online). 
Schematic flows of the right edges of LYZs as $N_{\rm max}$ and $N_{\sin}$ are varied. 
}
\vspace{-3mm}
\label{Nsamp}
\end{center}
\end{figure}

We also show the volume dependence of LYZs in Fig.~\ref{Vdep}. 
The imaginary number density in Fig.~\ref{thetadep} is calculated in the NJL model with infinite volume. 
However, the finite volume effect of the system appears from Eq.~\eqref{GCint} in the canonical approach. 
Following Refs.~\cite{Nakamura:2013ska,Wakayama:2018wkc}, 
we choose $N_{\rm max}$ so that the value of $N_{\rm max}/V$ is approximately unchanged when $V$ changes. 
Figure~\ref{Vdep} indicates that the right edges of LYZs, 
which is defined by a position of $\min[{\rm Im}(\xi)]$ in the first quadrant in the complex $\xi$ plane, 
approach the expected CP 
as $V$ increases. 
Note that we only calculate the right edge of LYZs for $VT^3=4^3$ due to a large computational time in the cBK algorithm. 
Because the right edges of LYZs are close to the expected CP at $VT^3\ge 2^3$, 
we can safely discuss extraction of the expected CP from LYZs at $VT^3= 2^3$.

\subsection{\label{xi_Nmax} $N_{\rm max}$ and $N_{\sin}$ dependences and fitting curves}

In the canonical approach, there are two parameters, $N_{\rm max}$ and $N_{\sin}$. 
In Fig.~\ref{Nsindep}, we show how the right edge depends on these parameters. 
We find that for finite $N_{\sin}$ as $N_{\rm max}$ increases, 
the right edges of LYZs pass over the expected CP 
and go to the origin. 
The behavior is more noticeable for smaller $N_{\sin}$. 
The reason why the right edges do not approach the CP but do the origin in large-$N_{\rm max}$ limit is that 
the imaginary number density is approximated by a Fourier series with finite terms as in Eq.~\eqref{mulsin}. 
In this case, phase transitions do not occur in the real chemical potential. 
Schematic flows of the right edges of LYZs against $N_{\rm max}$ and $N_{\sin}$ changes are summarized in Fig.~\ref{Nsamp}.

\begin{figure}
\begin{center}
\includegraphics[scale=0.70]{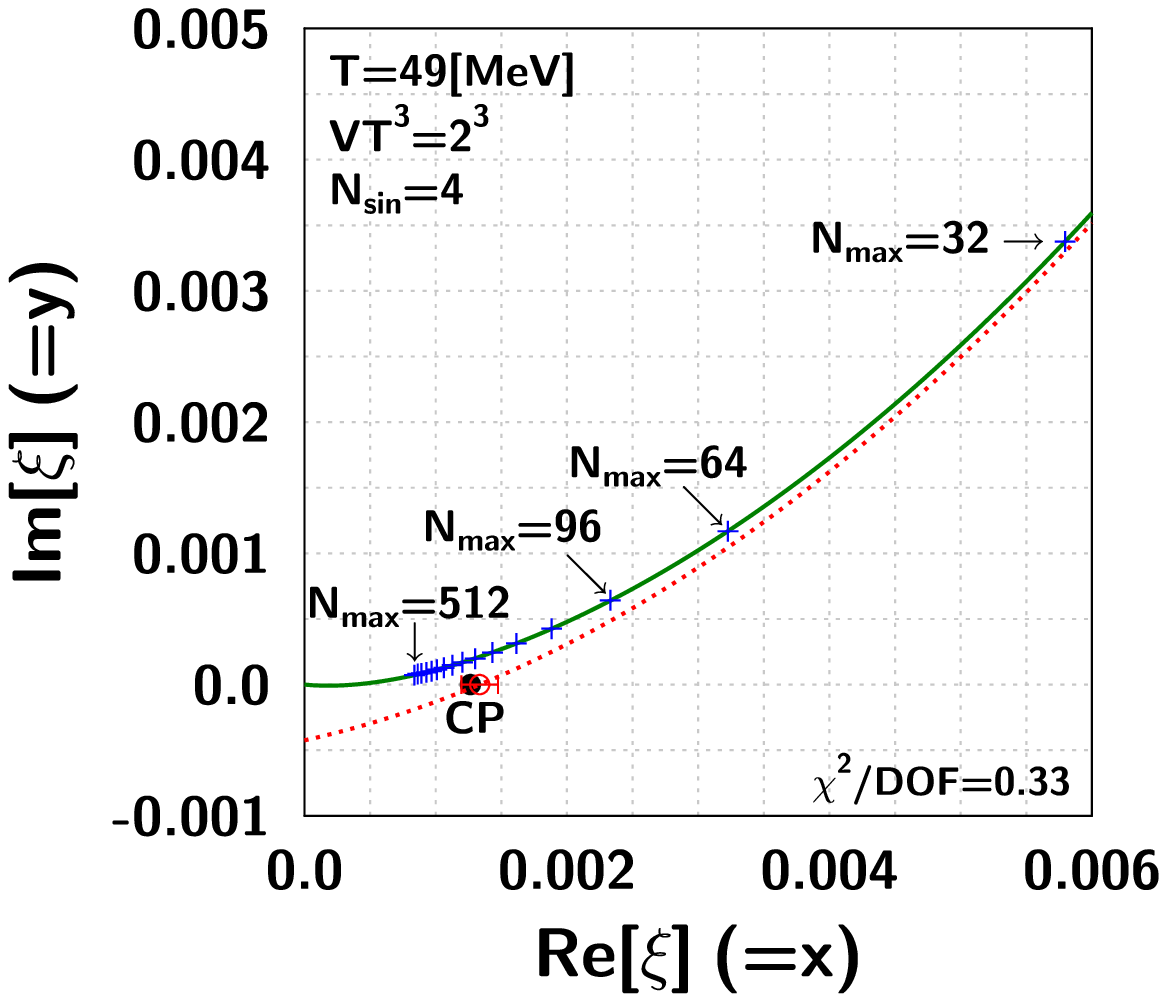}
\caption{ (color online). 
The $N_{\rm max}$ dependence of LYZs at $T=49$ [MeV] in the complex $\xi$ plane. 
The blue crosses stand for the right edges of LYZs. 
The solid and dotted curves represent the fitting function of the right edges of LYZs and the curve subtracting the first term from the fitting function. 
The red open circle means a point, $\xi=(x_0,0)$, where is the intersection of the dotted curve and the real axis. 
The black filled circle corresponds to the expected critical point (CP) calculated at the real finite chemical potential. 
}
\label{edgedep49}
\end{center}
\end{figure}

To obtain more insight on the results and to extract the phase transition point by the data with finite $N_{\rm max}$ and $N_{\sin}$, 
we try to fit the right edges of LYZs by a trial function. 
Figure~\ref{edgedep49} shows the $N_{\rm max}$ dependence of the right edges of LYZs. 
The points are computed at $N_{\rm max} = 32\mathchar`-$512 by a step 32 as shown there.  
We chose a fitting function as 
\begin{eqnarray}
\!\!\!\!\!\!\! y &=& \frac{b\br{cx_0-dx_0^2}}{x+b}+c\br{x-x_0}+d\br{x-x_0}^2 \ , \ \ \ \ \ \label{fit}
\end{eqnarray}
where $y={\rm Im}[\xi]$, $x={\rm Re}[\xi]$ with the parameters $b$, $c$, $d$ and $x_0$. 
This functional form is inspired by the expected behavior 
in the presence of the phase transition satisfying $y(x_0)=0$ in the limit $N_{\rm max}, N_{\sin} \to \infty$. 
This is implemented by the polynomial term which is nothing but the Taylor expansion at $x_0$, and hence $y(x_0) = 0$ is automatically satisfied. 
In contrast, for finite 
$N_{\sin}$, $y$ deviates from that and should approach the origin in the limit $N_{\rm max} \to \infty$. 
Indeed by adding the first term the trial functions satisfies $y(0) = 0$. 
In the absence of the phase transition, this argument can not be applied but is replaced by an instability of the fitting as we will discuss shortly. 

In Fig.~\ref{edgedep49}, 
the right edges of LYZs computed at $N_{\sin}=4$ and at various $N_{\rm max}$ from 32 to 512 with a step 32 are shown by blue cross points.  
The green solid line is the fitted function by Eq.~\eqref{fit} and the red dotted line the one of the polynomial term. 
The fitting is performed by the weighted least-squares method 
where the weights $1/(\delta y_i)^2$ are given by the systematic errors of the cBK algorithm $\delta y_i$. 
We find that 
the data points of the right edges of the LYZs are reproduced with good accuracy: $\chi^2/{\rm DOF}=3.95/12$. 
As if to prove that the first term is the finite $N_{\sin}$ effect, the curve subtracting the first term, 
which is represented by the dotted curve, 
crosses the real axis at $x_0=1.33(14)\times 10^{-3}$. The result is consistent with the value of expected CP, $\xi_{\rm CP}=1.26\times 10^{-3}$. 
Note that the other parameters $c$ and $d$ in the resulting curve are also the same as the ones of the fitting function: 
$c=4.15(34)\times 10^{-1}$ and $d=7.26(15)\times 10$.

\begin{figure*}
\begin{center}
\includegraphics[scale=0.70]{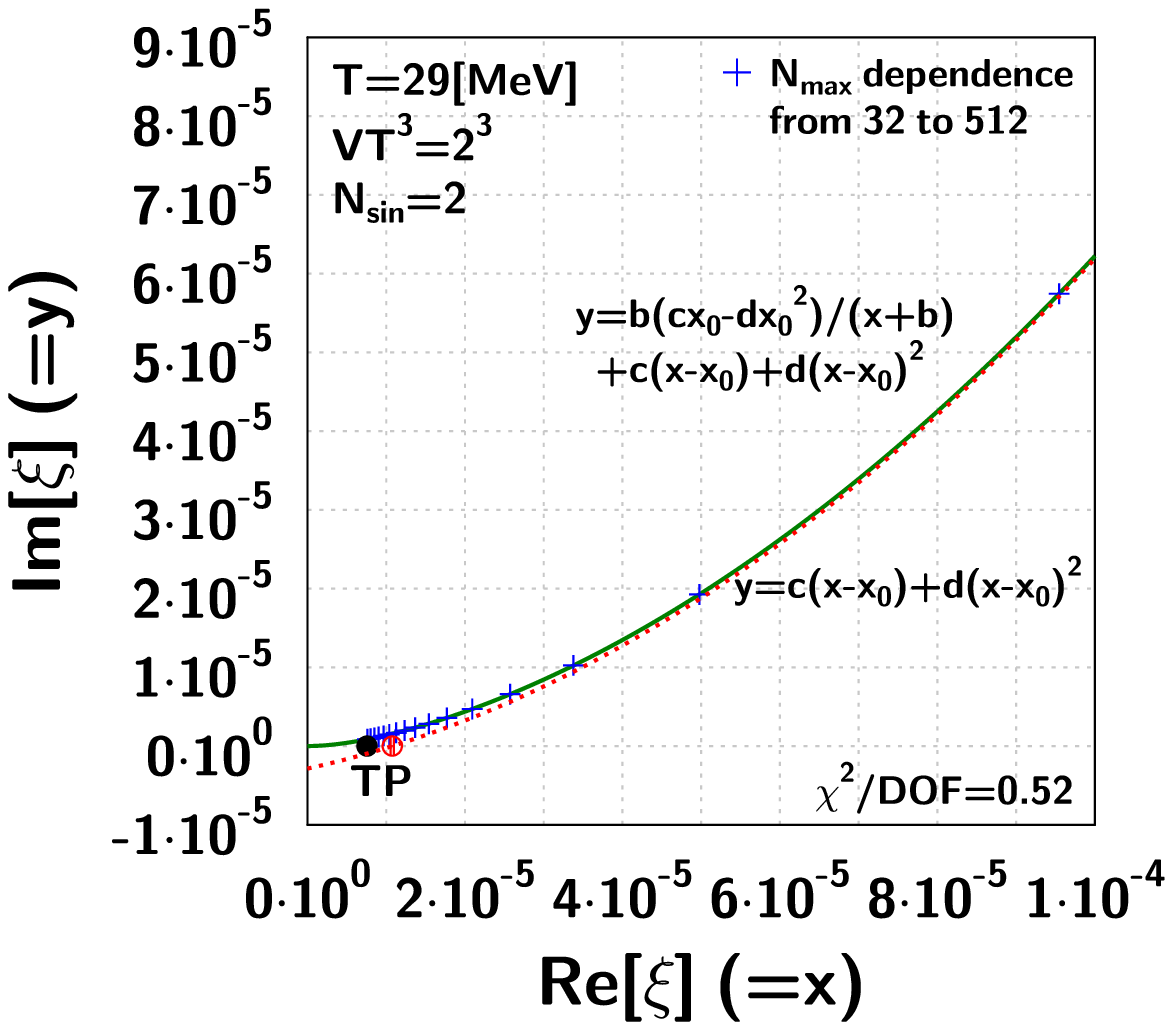}
\includegraphics[scale=0.70]{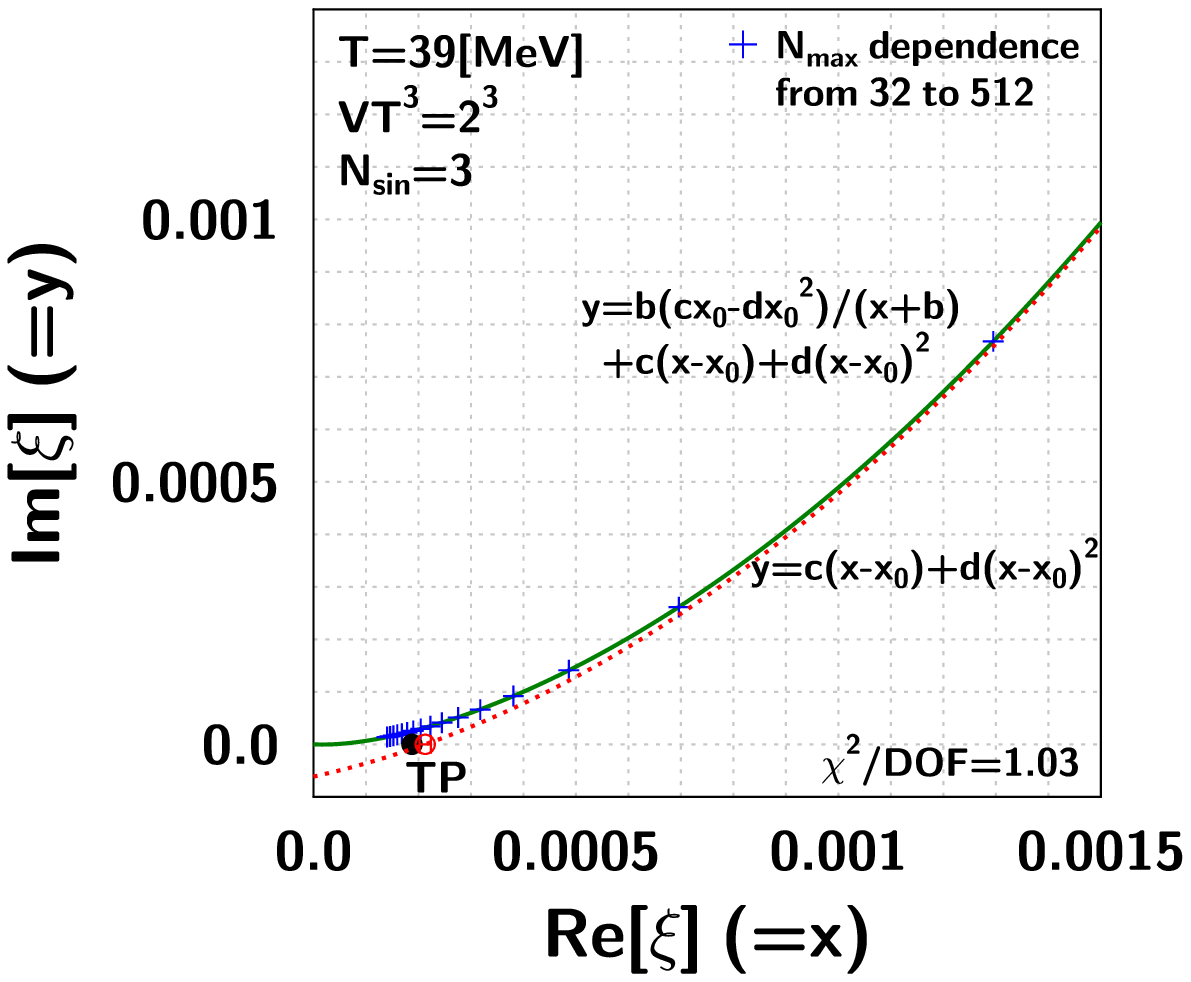}
\caption{ (color online). 
The $N_{\rm max}$ dependence of LYZs at $T=29$ and 39 [MeV] in the complex $\xi$ plane. 
The blue crosses stand for the right edges of LYZs. 
The black filled circle corresponds to the expected transition point (TP) calculated at the real finite chemical potential. 
}
\label{edgedep29}
\end{center}
\end{figure*}
\begin{figure*}
\begin{center}
\includegraphics[scale=0.70]{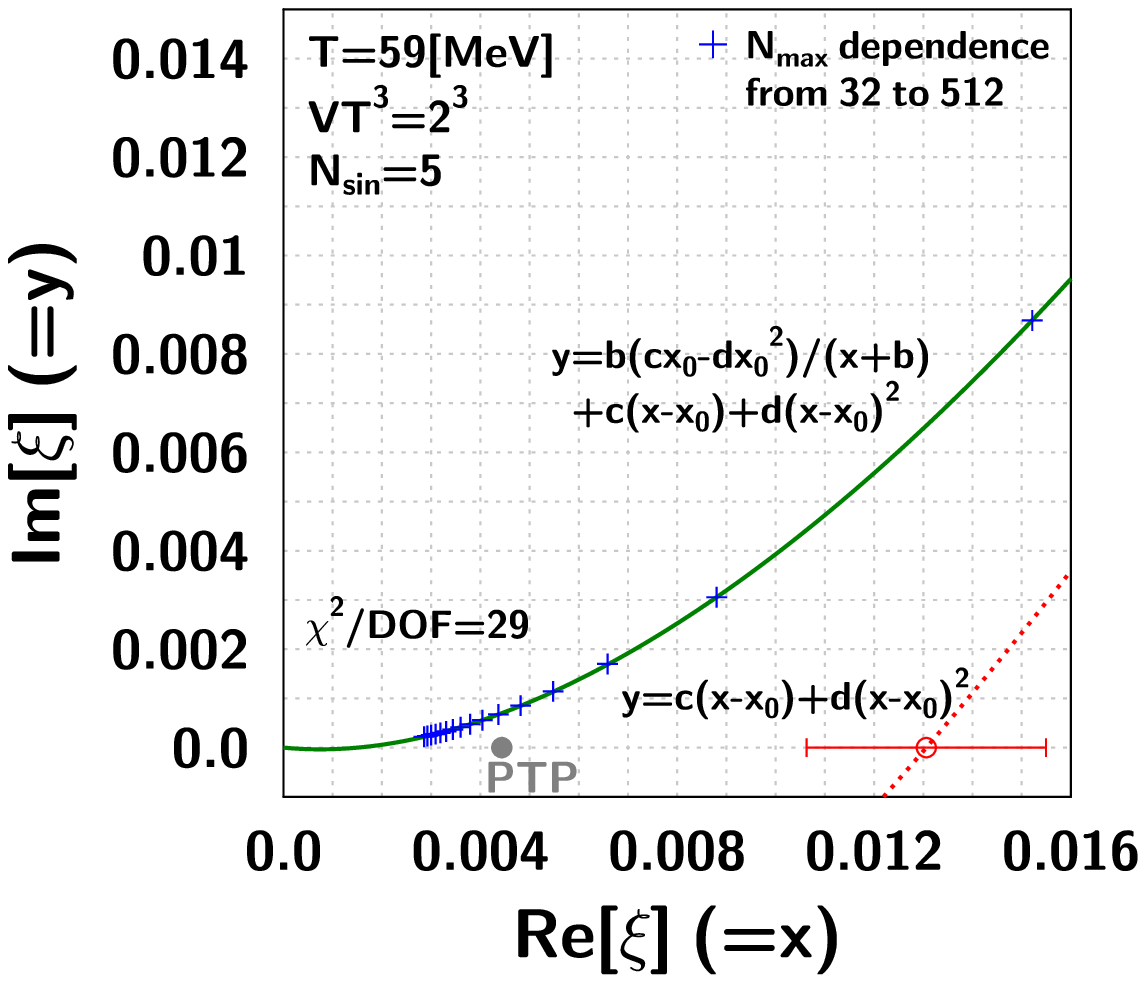}
\includegraphics[scale=0.70]{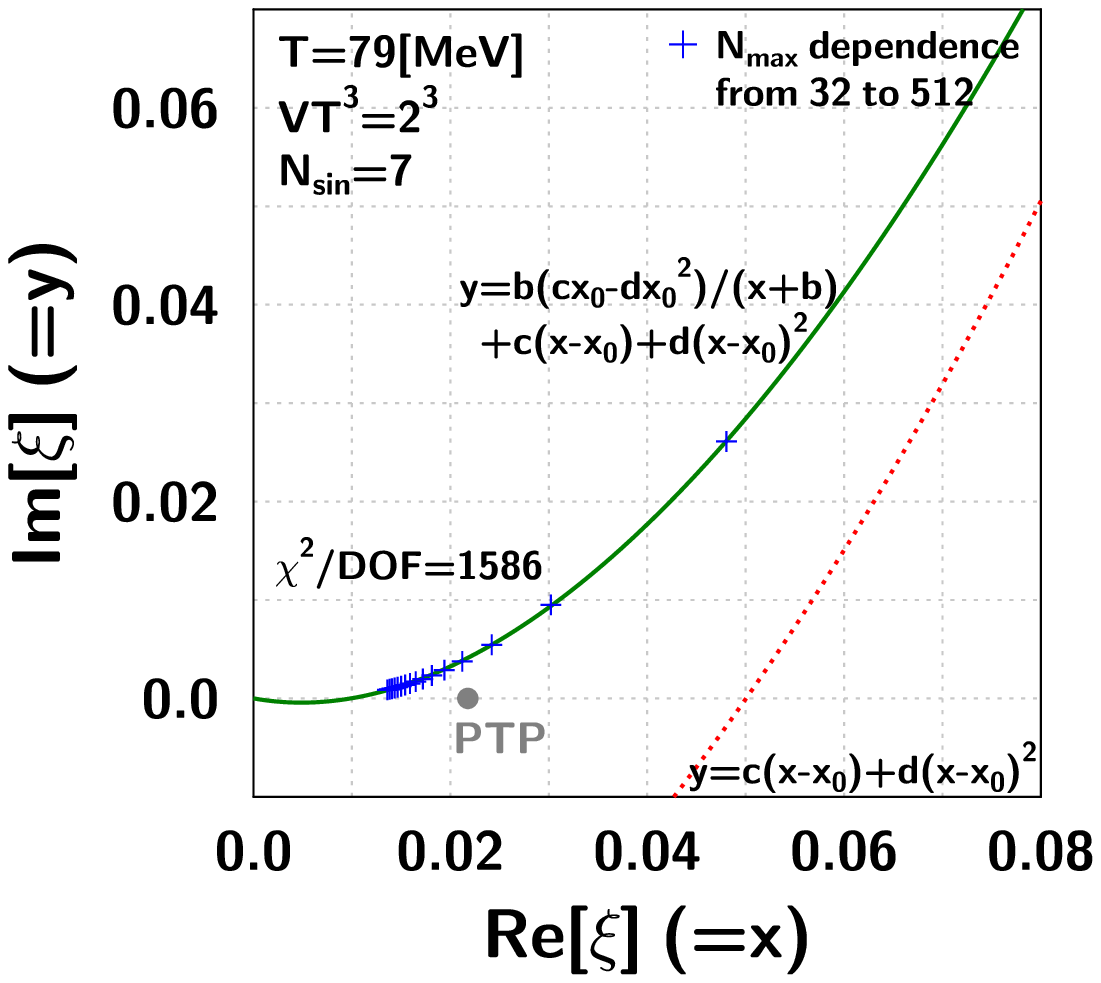}
\caption{ (color online). 
The $N_{\rm max}$ dependence of LYZs at $T=59$ and 79 [MeV] in the complex $\xi$ plane. 
The blue crosses stand for the right edges of LYZs. 
The gray filled circle corresponds to the expected pseudo transition point (PTP) calculated at the real finite chemical potential. 
In the right figure ($T=79$ [MeV]), the intersection of the dotted curve and the real axis is not represented 
because the intersection has a large error, $x_0=0.050(137)$. 
}
\label{edgedep59}
\end{center}
\end{figure*}

In Fig.~\ref{edgedep29}, we also show the $N_{\rm max}$ dependence of LYZs at low temperatures, 29 and 39~[MeV]. 
Then, we find that the extrapolation procedure performed in the simulation at $T=49$~[MeV] works well to obtain the expected transition points (TP) 
with good accuracy. 
Therefore, we can conclude that the extrapolation procedure of LYZs is reasonable to extract the correct transition points from the canonical approach.

Finally, Figure~\ref{edgedep59} shows the $N_{\rm max}$ dependence at high temperatures, 59 and 79~[MeV] 
where there is no phase transition. 
It is interesting to see how the extrapolation procedure works in this case. 
From values of $\chi^2/{\rm DOF}=29$ ($T=59$ [MeV]) and 1586  ($T=79$ [MeV]), 
we find that 
the fits of the LYZs become unstable at high temperatures. 
In Fig.~\ref{edgedep59} we also show 
the peaks of susceptibility of the number density at the crossover represented as the pseudo transition points (PTP). 
Our results obtained from the extrapolation procedure are far from the PTP, 
which is consistent with the disappearance of phase transition points. 
The results indicate that 
the stability of the fit of the right edges of LYZs distinguishes between regions with and without phase transition points.

\section{\label{summ}Summary}

We have investigated phase transition points from LYZs calculated in the canonical approach of the two-flavor NJL model. 
After we have extracted the canonical partition functions from the imaginary number density 
by using the integration method and multiple-precision arithmetic, 
the LYZs have been evaluated with the cBK algorithm. 
In the integration method, we have fitted a Fourier series to the 161 data of 
the number densities as functions of imaginary chemical potential. 
Because phase transition structures of the NJL model are already known 
with the critical point, $(T_{\rm CP},\mu_{\rm CP})\simeq (49,327)$ [MeV], 
simulations have been carried out at several temperatures around $T_{\rm CP}$, 
$T=29$, 39, 49, 59 and 79 [MeV]. 

We have shown how the LYZs behave as functions of the volume of the system $V$, $N_{\sin}$ and $N_{\rm max}$. 
The results of the $V$ dependence at $T=49$ [MeV] have shown that 
the simulation $VT^3=2^3$ has a sufficiently large $V$. 

In the investigation of the $N_{\rm max}$ dependence of LYZs, 
we have extrapolated the right edges of LYZs from finite $N_{\rm max}$ to the infinite $N_{\rm max}$. 
We have succeeded in subtracting a term associated with an artifact due to finite $N_{\sin}$ from the fitted function at $T \le T_{\rm CP}$. 
We have found that the curve 
with the finite $N_{\sin}$ effect subtracted 
crosses the real axis of $\xi$ near the expected transition point calculated at the real finite chemical potential. 
The extrapolation procedure becomes unstable 
at $T > T_{\rm CP}$, 
which is consistent with the lack of phase transition points at the real chemical potential. 
The results indicate that 
the accuracy of the fitting of the right edges of LYZs distinguishes between regions with and without phase transition points. 

In this paper, we have discussed LYZs of the NJL model and have found a reasonable extrapolation procedure. 
Whether the extrapolation procedure has universality for models or not is open to discussion.

\section*{Acknowledgments}
This work was supported by the National Research Foundation of Korea (NRF) 
grant funded by the Korea government (MSIT) (No.~2018R1A5A1025563). 
AH is supported in part by Grants-in-Aid for Scientific Research (No.~JP17K05441 (C)) and for Scientific Research on Innovative Areas (No.~18H05407). 
This work was supported by 
``Joint Usage/Research Center for Interdisciplinary Large-scale Information Infrastructures" and 
``High Performance Computing Infrastructure" in Japan (Project ID: EX18705 and jh190051-NAH). 
The calculations were carried out on OCTOPUS at RCNP/CMC of Osaka University.



\end{document}